# Distinct 3D Glyphs with Data Layering for Highly Dense Multivariate Data Plots

Santiago V. Lombeyda, *California Institute of Technology*

**Abstract**—A carefully constructed scatterplot can reveal plenty about an underlying data set. However, in most cases visually mining and understanding a large multivariate data set requires more finesse, and greater level of interactivity to really grasp the full spectrum of the information being presented. We present a paradigm for glyph design and use in the creation of single plots presenting multiple variables of information. We center our design on two key concepts. The first concept is that visually it is easier to discriminate between completely distinct shapes rather than subtly different ones, specially when partially occluded. The second one is that users ingest information in layers, i.e. in an order of visual relevance. Using this paradigm, we present complex data as binned into desired and relevant discrete categories. We show results in the areas of high energy physics and security, displaying over 6 distinct data variables in each single plot, yielding a clear, highly readable, and effective visualization.

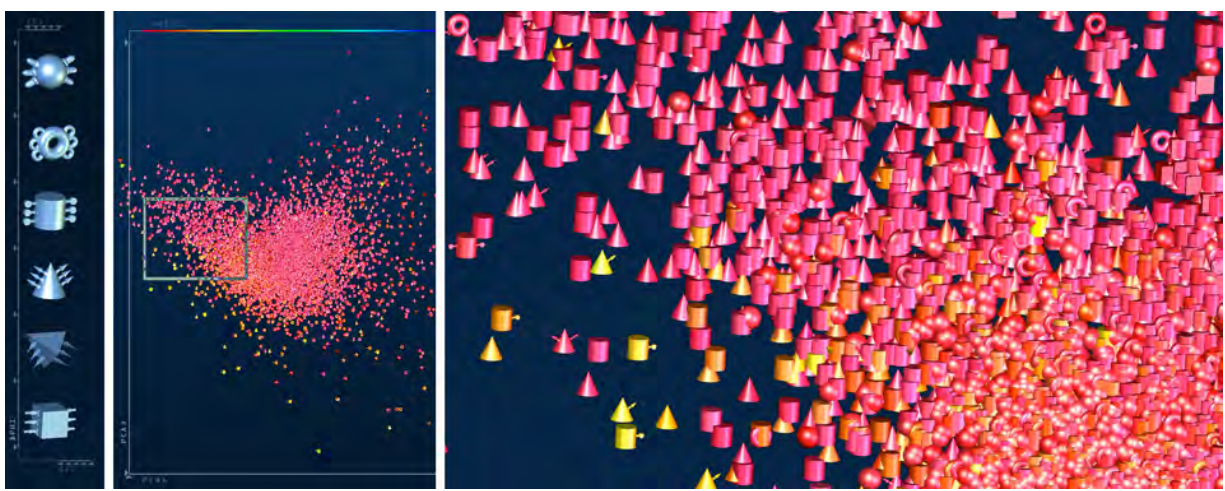

Fig. 1. (Left) Six unique 3D glyphs used to map the angle between two particle tracks (`dphi`) from a high energy physics detector simulation. Left "arm" features on glyph map energy from one photon (`eg1`), right "arms map energy from second photon (`eg2`). (middle) Scatterplot mapping two principal components to (x,y) position. Glyph hue represents Higgs Mass. (Right) Close up of left branch of scatterplot. Lower-right Dense plot area is easily recognizable as mostly spheres (low angle) but for few tori and cylinders. Upper-left area shows mostly cylinders and cones. Few objects show high energy as expected to correlate to mass.

## 1 INTRODUCTION

The effectiveness of a visualization paradigm ultimately depends on its ability to enable the particular visual mining goal a researcher is trying to accomplish.

In many real data analysis situations, there may be an unclear expectation of the location and form, and even the process needed to reach a discovery from visually mining the data. In such cases, presenting as much information to the researcher as possible may be an advantageous first step. However, in the process of doing so, it is important to ensure that the information presented remeains clear, readable, and enables well the discovery process. The challenge with multivariate data is not just in showing multiple variables at the same time, but being capable of showing the relationship between them without overwhelming the eye.

A scatter plot matrix [4] is a very successful paradigm at showing how pairs of variables relate. However, in order to correlate one single scatter plot – i.e. a pair of variables – to additional variables, follows a need for further interactive techniques. Such tools include brushing, which allow to cross-reference information visually by marking subsets of points across all scatter-plots with a brushed-on color.

Meanwhile, parallel coordinates [3] combined with brushing, sub-setting, and other interactive techniques, make it easy to see how selected data records map across all plotted variables. However parallel coordinates lacks the ability to well depict physical and spatial relationships that map well to scatter plots.

Overall, systems that do tie these analysis tools together interactively and offer supplemental statistical analysis mechanisms, allow these paradigms to complement each other and can result in a powerful data analysis tool.

However, optimally, a researcher would be shown a single plot that by itself conveys valuable information needed to visually segment the data, find clusters, find patterns, and find outliers. This would be particularly helpful, and needed, given a large number of data points (>10,000), with multiple variables per data entry (>6). So, as we know already, in order to expose complex data relationships, it is necessary to simultaneously correlate as many

---

- *Santiago V. Lombeyda is with the Center for Advanced Computing Research at the California Institute of Technology, E-Mail: slombey@caltech.edu.*

meaningful variables as needed. Thus to show these in one single graph it would entail "layering" all additional vital information on "top" of a basic plot.

Our proposition solely focuses on increasing the amount of information shown on a single plot, with special attention to readability and clarity. We achieve this by binning the data and mapping it to discrete visual attributes tailored to exhibit pre-designed readability order.

## 2 CONTRIBUTIONS

There is a trade-off between clutter and the visual uniqueness and amount of information each glyph can represent. For instance, [1] faces can represent 6+ variables. However, recognizing patterns or clusters from the details of these faces is not an easy task.

We sustain there are two corroborating reasons for this. When the change between glyphs is too gradual, it is easier to loose the ability to detect clusters, patterns, or outliers through unclear glyph features. This is a well-known phenomenon that presents itself often in medical image analysis, when using color transfer functions with only smooth low frequency changes. One solution is to increase the complexity of the color transfer function, by adding higher changes in color and intensity. The other alternative is to mine the data and figure out where the significant changes occur, and then map those to a correspondingly adjusted color transfer function. Notice that in the first choice, we rely on the user visual perception skills to understand the resulting images; assuming the generated function is able to pick up the important details. The second choice, while seemingly more comprehensive, is much harder to implement, is more compute intensive, and may ultimately still miss details, that a trained human can still easily recognize.

The second reason we support that complex glyphs like Chernoff faces have not succeeded, relates more to the visual design and geometry of glyph itself. Most importantly missing, is "reading order". It is well understood that when there is lack of a clear hierarchy of the visual importance of elements [9] of an image, a viewer's eyes try to distribute attention between the competing elements, which leads to eye tiredness. This simply translates to a feeling of clutter and confusion. The solution is to have a clear layering, or order of visual importance of the information. In such way they viewer's first glance is clear, and then can choose to visually dive (i.e. focus) into the second and following layers of information as needed.

## 3 PREVIOUS WORK

The ability to present multiple variables of data simultaneously on a single image, and accurately portray the relationships between the variables is one of the core challenges in scientific visualization. Given a good grasp on the meaning of the data and its intra-relationships, and with a strong sense of design, visualizations can be extremely successful, as those often quoted by artists, designers, and scientists, the likes of the work of Richard Tufte [15]. However, when trying to construct a more general tool, our best option is to present the maximum amount of data. While doing so, we need to preserve the same successful visual design attributes that will present the information in the most understandable and meaningful fashion.

The 1973 work of Herman Chernoff [1] introduced the use of faces as glyphs, where the attributes of the faces (head eccentricity, eye eccentricity, pupil size, eyebrow slant, nose size, mouth shape, eye spacing, eye size, mouth length, and degree of mouth opening) were used to encode data. The assumption was that the human eye is well trained in face recognition and is quite keen on recognizing facial expressions. While it is possible that with more modern 3D rendering techniques we could re-test its proposition on realistic looking faces, overall it has been tested that Chernoff faces are not particularly perceptually efficient [12]. Furthermore, [7] showed that when plotting multivariate data, subtle glyph differences are hard to detect or evaluate.

The work of [6] presents tensor data as oriented ellipsoids. Orientation as well as ellipsoids radiuses present the 6 variables of the reduced tensor. Meanwhile, their spatial arrangement on a 2D grid (as based on the underlying scanned images) creates a sense of texture, similar to that of art, such as the work of the late 19th century impressionist master, Vincent Van Gogh. However, this work does not translate well to unstructured data, as possible overlap between the glyphs would be confusing.

It is in fact this same issue with complex glyphs such as Chernoff faces that we can argue about the use of most complex glyphs, including radar plots (whisker stars) and even 3D glyphs based on superquadrics [13]. While unique, and given enough time to analyze one glyph does convey all the n-variable information for each data record, it is mostly the large eccentricities that are perceived on complex plots. But most importantly, when overlaying each other, regular radar graphs can create clutter or be altogether undistinguishable.

Truly unique icons were proposed in [8] for file browsers. Their central proposition was that content-meaningful icons are not as valuable as easily recognizable icons solely in the task of facilitating users to find their documents. They developed a mechanism to generate unique icons, purely based on the filename. Files with similar filename, would have similar, but still different, icons. Their icon generation was based on hash-function based recursive calls to basic shape generators. Their icons were mostly line illustration based, though were most aesthetically pleasing, when engulfed in glass like bubbles, better defining the icon boundary and engulfing the created shape. Through repeated use, users became familiar with each individual icon, regardless of its innate visual content. This work however cannot be translated to a visualization plotting system, as these complex constructions are not amiable to be overlapped while remaining discernable from each other.

## 4 3D GLYPHS

Figure 2 shows a unique small set of glyphs that have been crafted to be recognizable even in highly dense plots. Our goal is for these glyphs or icons to encode several variables of information for each data record they represent, while still being easily visually differentiable. We propose to encode each additional layer of information as corresponding additional features on the glyph with a decreasing level of visual importance.

In order achieve our clarity goals we set forth the following requirements.

- *Dimensions*: We use three-dimensional glyphs so as to allow partial occlusion and partial extrusion specially when glyphs resting on the same plane.

- *Size*: We set all glyphs to be the same perceived size.

- *Shape*: We choose glyphs that when partially occluded are still mostly distinct from the other glyphs

- *Orientation*: We set glyphs to always be facing the same direction.

### 4.1 On Dimensions

The use of 3D glyphs is crucial, particularly for 2D scatter plots, where all icons rest on the same plane. The use of 3D glyphs allows for several icons to be in extremely close proximity while some characteristic regions of the shape to protrude --assuming good glyph design.

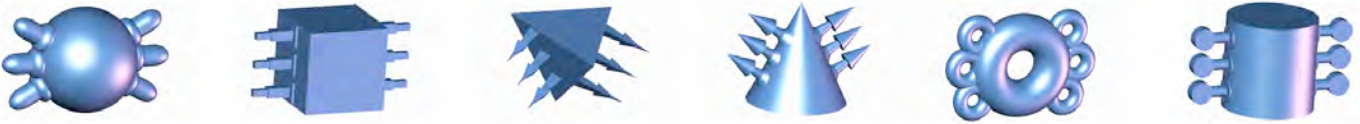

Fig. 2. 3D Glyphs, with full left side and full right side data features.

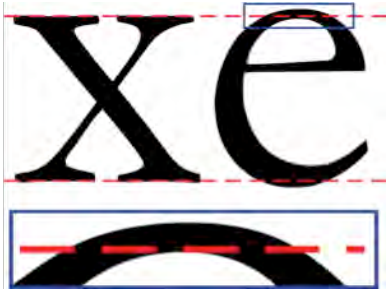

Fig. 3. Classic example of typographic design, where curved shapes (letters) are drawn beyond the guide lines (*base line* and *x-height* overhang) to allow letters to seem visually equivalent in height. Shown font is Adobe's Garamond.

When flat 2D glyphs, as used on classical plotting systems, are overlapped the resulting image varies depending on the rendering order. Rendering order becomes unimportant for overlapping (non-coincident) data points when 3D glyphs are used.

Because the requirements we impose on size and orientation, 3D rendering even of large number of glyphs can be easily managed. "Pre-backed" renderings of (colorless) glyphs can be made with all permutations of the different discrete additional characteristics. These can be then stored as sprites with depth (i.e. z-buffer), and then used to quickly render highly complex scenes. This technique is particularly useful in systems with limited graphic processing capabilities, which includes most compute clusters.

### 4.2 On Size

In order for glyphs to appear to be the same size, we make glyphs with a curved profile slightly larger than those with flat profiles. This is a well-known visual effect that is commonly addresses in typography in the same fashion. See figure 2.

We extend this approach to sharp objects, like the cone. We take similar considerations of the sides of the glyph, so as to appear of similar width; as well as protrusions from the front face of the object (towards the viewer) that will aid in the object being recognizable. We address all this issues manually upon design of the glyphs, and use visual inspection.

### 4.3 On Shape

The construction of these glyphs was done with several criteria in mind: front profile, top profile, protrusions, and corners versus curved regions.

For the front profile, we started with most basic shapes that are common to 2D plots. These include circles, triangles, stars, boxes, pluses and exes. We proceeded to derive 3D shapes with these listed front profiles. Circles translate to spheres and tori. Triangles translate to pyramids (or tetrahedral) and cones. Boxes translate to cubes and cylinders. While exes and pluses could be made intro 3D glyph in a similar fashion, each branch of the glyph has a potential to extrude beyond an occluding object and be confused for a different type of glyph or a feature in a simpler glyph. Same difficulty as can be encountered with star plot based 3D icons. It should be possible to very carefully construct a 3D icon from these that is unique and does not lead to these possible visual confusing situations. We believe the number of glyphs presented in this work is sufficient to be efficient yet perceptually manageable.

These chosen basic shapes demonstrate a variety of top and bottom profiles: circle, square, point, triangle, and ellipse. So, while some glyphs do share same front profile, particularly the cylinder and cube, the number of protruding corners (or lack there off) and top profiles allow them to remain recognizable even on densely populated plot regions. Furthermore, the lighting exacerbates the visual distinction between shapes, aiding in the ability to differentiate between them. Similarly, the cone and pyramid, depending on orientation, can exhibit same front profiles, i.e. a triangle. We thus chose the cone to point upward, while the pyramid was set to have one horizontal edge at the top, while pointing outwards.

### 4.4 On Rotation and Orientation

If we were to design a glyph where its faces were parallel to the front plane, then when overlapped with same others, they would create a large plane, rather than a more complex surface we desire. Likewise, if the top and bottom profile were aligned with the floor, when using an orthographic projection --as we mostly do-- the top and bottom profile would be hidden. We therefore rotated all icons the same amount both horizontally and vertically. The resulting rotation is equivalent to a 30-degree rotation around an axis 30-degrees west of north (i.e. 30 degrees right of screen up vector). As we will discuss later, the addition of extra features on the side of the object will also benefit from the horizontal rotation component, in order to give more visual importance (via proximity) to one side over the other.

Meanwhile, as pointed out repeatedly, the glyphs are carefully crafted to have unique appearances, especially when highly occluded. Thus, when extending the use of these glyphs to a 3D space, as 3D graph, having the glyphs preserve their initial orientation both enforces their appearance and their designed ability to be recognized when overlapping.

| | | | | | **OCCLUDING SHAPES** | |
|---|---|---|---|---|---|---|
| | balls | boxes | tets | cones | tori | cans |
| ball | 5% | 48% | 92% | 91% | 42% | 53% |
| box | 47% | 4% | 86% | 92% | 58% | 38% |
| tet | 25% | 27% | 6% | 69% | 44% | 18% |
| cone | 22% | 12% | 66% | 8% | 28% | 10% |
| torus | 65% | 19% | 93% | 83% | 8% | 49% |
| Can | 44% | 51% | 92% | 87% | 57% | 5% |

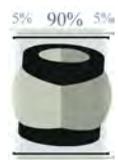

Table. 1. Occlusion tests results for 90% overlap between test object and 2 occluders on each side. Measured in percent of visible pixels.

### 4.5 Glyph Shape Evaluation

The foremost evaluation that has to take place is a visual inspection of the shapes resulting from each glyph when it is partially occluded. We do so by constructing a grid of each glyph overlapped on two sides by each of the other glyphs, as presented through an occlusion matrix as in figure 5. Purely from visual evaluation it is easy to pick out glyphs that may violate any of the premises we have established.

Beyond visual evaluation, we designed an empirical occlusion test that evaluates the number of pixels visible for a glyph as partially occluded by other two. We perform these tests at a range of different overlapping distances. The results helped in validating our design premises. For all shown glyphs, when bounding boxes overlapped on 90% of the center region of the test object by both "occluding" objects, and 5% on each side of the test object by just one occluding object; on average 54% of possible pixels were still visible. This excludes same shape comparisons between occluder and test object. When the occluding shape was the same as the shape being tested, the average of theses cases was 6% of possible pixels visible, with the box being the worst case at 4%, and cone the best at 8%. As can be see in table 1, the box glyph is the worst (i.e. most efficient) occluder, while the sphere glyph is the best at protruding from occluders.

Overall, the reason for these high results is due to the careful design of the shapes. If the shapes were not given the slight rotation then the boxes may completely occlude all other objects. Similarly, setting the objects to be visually equivalent in size, means that objects like the cone and pyramid have extrusions that stick out of even highly occluding objects like the cylinder and the box.

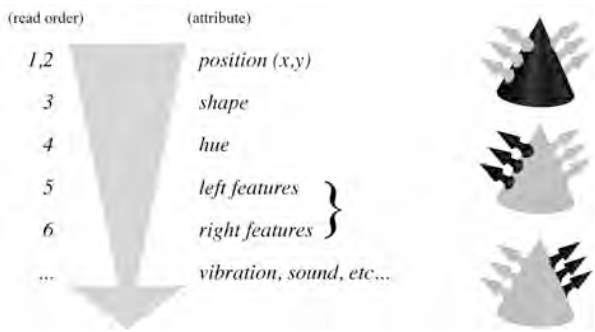

Fig. 4. (Left) Desired visual importance order goal design. Includes possible future extensions (Right) Left side, and right side features.

In the end, it is not only necessary for the glyphs to be capable of being visible when highly occluded, but it is the uniqueness and quality of what pixels are visible that will dictate good glyph design. Notice that similar physical qualities between a sphere and a dodecahedron, or between a cone and a regular tetrahedron oriented exactly the same, make them bad candidates to be both used on the same plot. Ultimately, the way to determine this is through thorough visual inspection.

## 5    Data Layering

When encountering a scatter data-plot, the first visual impression is that of the overall spatial data distribution. Upon further inspection the visual qualities of the data points become relevant. From color research [4] we know that luminosity is more important visually than hue. Our use of 3D glyphs results in an image where the shapes of the objects are mostly encoded across the luminosity channel due to the standard lighting model and basic scene environment setup. Thus, color (i.e. hue) will be subordinate to shape.

Given our choice for standardized size glyphs with piecewise smooth shapes, we can encode further information as additional shape features, or "arms". We define arms as geometric protrusions, orthogonal to the surface, constructed by smaller versions of the original glyph, and in the same color. We limit the number of arms to three, to each the left and right side of the glyph. In order to establish a difference between encode binned variables on each side, we enforce a slight tilt of the object towards the right. This forces the arms on the left side to appear more prominently than the arms on the right side. Each variable is binned into four buckets, from which they are mapped into 0 to 3 arms. While we could have device a binary permutation scheme to increase the number of buckets encoded, it would have unnecessarily increase the glyphs complexity.

Finally, we use straight binning of the data by dividing the variable's domain into even intervals. If the data is mostly concentrated into one bin, that is something that will become easily apparent through our paradigm. However, we do offer three alternatives to this situation. First, we can iterate a desired number of times, re-dividing overly populated bins, in order to achieve better load balance. Second alternative is to offer interactive sliders to allow for varying limits of each of the bins. Finally, we can also do a simple analysis of each bin, and define a new variable for the distribution in each bin, which can be mapped to a different attribute. This is somewhat equivalent to separating a signals low frequencies from the high ones, and visualizing them separately.

## 6    Results and Conclusion

We present two sample utilizations of our proposed 3D glyphs. We collaborated with experts in both fields to ensure that the results were both accurate, useful, and offered novel insight into their data.

We believe that our results are quite successful, and can be coupled into modern data exploration systems, to offer greater power for researchers to do visual data analysis. We offer a set of glyphs that can be empirically tested for their ability to be perceivable even on highly dense plots, and show how to visually test their effectiveness through uniqueness. We also present a way to layer data, and we show how having an a priori goal for the way the layers of information are perceived can help achieve meaningful multivariate plots. Finally we show highly dense plots, where not only it is still possible to recognize glyphs even quite occluded; but the resulting image of these regions can be read as a bump map, or as a texture encoding patterns of the data represented.

### 6.1    Visualizing High Energy Physics Events

Current research in High Energy Physics has lead to groundbreaking and awe inspiring scientific projects such as CERN's Large Hadron Collider (LHC) detector. The amount of data that can be gathered from such a tool and how to analyze it is an open research goal scientists are trying to figure out. Scientists are currently generating simulations that mimic the data coming from the LHC, which have already been pre-classified as meaningful signal events versus background events. This distinction is something researchers will have to do from the actual data as it is gathered. As researchers dive into the current simulated data, they start by visually mining it in order to find patterns, possible way to cluster the events, ultimately classify them as signals or background events, and specially trying to understand outliers and their meaningfulness.

We begin our study by mapping the pre-classification between signal and background noise into a separate variable, and then doing principal component analysis (PCA) on the data (without the classification variable). We construct a correlation table to better understand how the created PCA components map to the original variables. We further create an outlier evaluation that creates a rank variable. For all of this we use a combination of Mondrian [11] (R based [14]) and Mollegro's Data Modeler [10].

From this analysis we learn that the energy for each pair of photos is fairly independent of other variables, as is the Higgs mass. Furthermore, the Higgs mass highly correlates to the outlier rank. We create a plot, where we map two principal components to the Cartesian coordinates, the delta Phi (angle between particle paths) we bin into 6 buckets corresponding to each shape, the Higgs mass we map to color, and each photon's energy we bin into four buckets and map it to arms on left and right respectively. See figure 1.

We analyzed the resulting visualization with a High Energy Physics field expert, and right away there were features of the overall plot that were quite meaningful. Few outliers, due to their spatial location were easy to detect. Due to the additional layered information it was also easy to learn more about the specifics of these data points, which was quite meaningful to the researcher. Notice in for instance the event represented by a green cylinder at the bottom, which was of particular interest to our expert. Both the color and the number of arms corresponded to a high energy, and high mass event, which he would have expected would be true, and was able to confirm. Finally, we created an alternative plot, where pre-classified signals were mapped closer in depth versus background noise events, creating a seven variable plot, which we can use to verify hypothesis gathered from the unclassified version.

### 6.2    Other Results

Figure 7 shows a subset of the 2008 VAST challenge dataset, as graph created by applying Eigen state reduction. We encode connections in the graphs with semi-transparent lines, to set them at less visual importance than the position and shape of the data points. The width of the line is maps to bins for call length. Thus this one plot presents nine meaningful variables.

## 7    Future Work

This paper focuses on the design of unique glyphs that meet our criteria for clear display of multivariate data in high density plots. We believe some further work is warranted both in layering extra information into our paradigm, as well as aiding in the exploration of this complex plots.

We are currently experimenting with adding subtle motion (jitter or twist) to the glyphs. We believe, that in keeping with our experience, binning an extra data variable, and mapping it to four different rates of motion/vibration (none, mild, medium, and fast) allow us to layer additional information, without compromising our goal of clarity. We however have not reached a conclusion as to how such motion is still acceptable or simply distracting, and whether layering extra motion on different axis is possible (or yaw, pitch, and roll), and differentiable from one another.

Additionally, while dense plots can create visual textures that encode valuable information, we complemented our work with the ability to explore these dense areas, and probe individual glyphs. We use an interactive localized magnification algorithm, to expand glyphs within a small area., as can be see in figure 6.

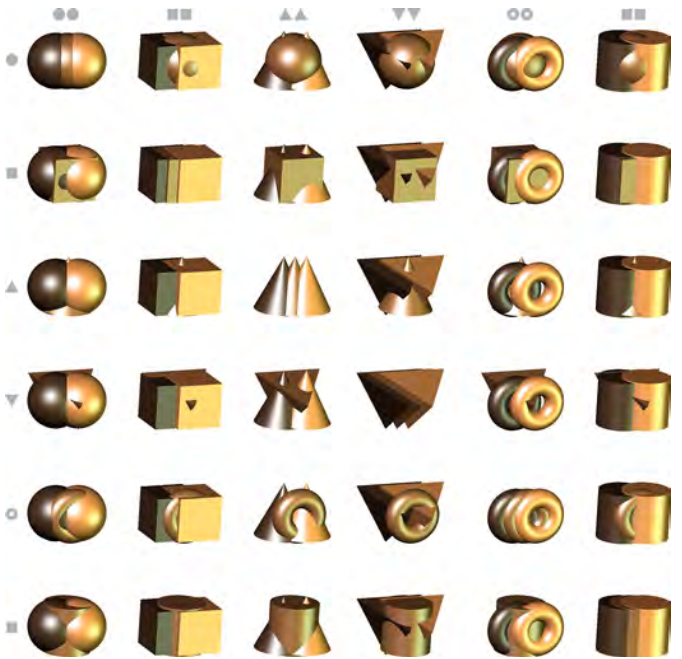

Fig. 5. Occlusion test matrix. Rows: test glyphs, Columns: occluders.

### ACKNOWLEDGEMENTS

We would like to thank: Julian Bunn for providing High Energy Physics data, and serving as an expert evaluator from the field; Roy Williams for providing the VAST graph and serving as expert evaluators from the field; Mathieu Desbrun for aid in refining of ideas and editing. This work was funded by the Moore Foundation through Caltech's Cell Center, and by the NNSA's Predictive Science Academic Alliance Program (PSAAP), through Caltech's PSAAP Center of Excellence.

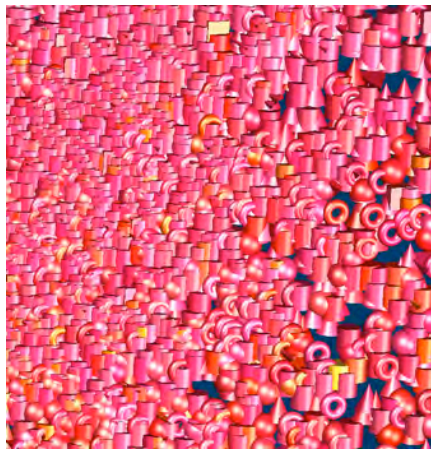
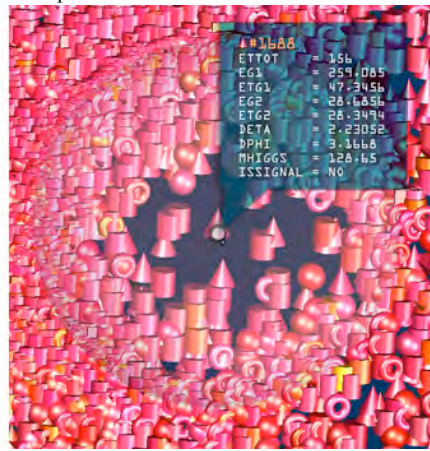

Fig. 6. Sample magnifying lens exploration of dense plot. Left image shows a highly dense plot region of HEP data set. Right shows how a simple transform, can expand the space between glyphs in region of interest, to better explore contents.

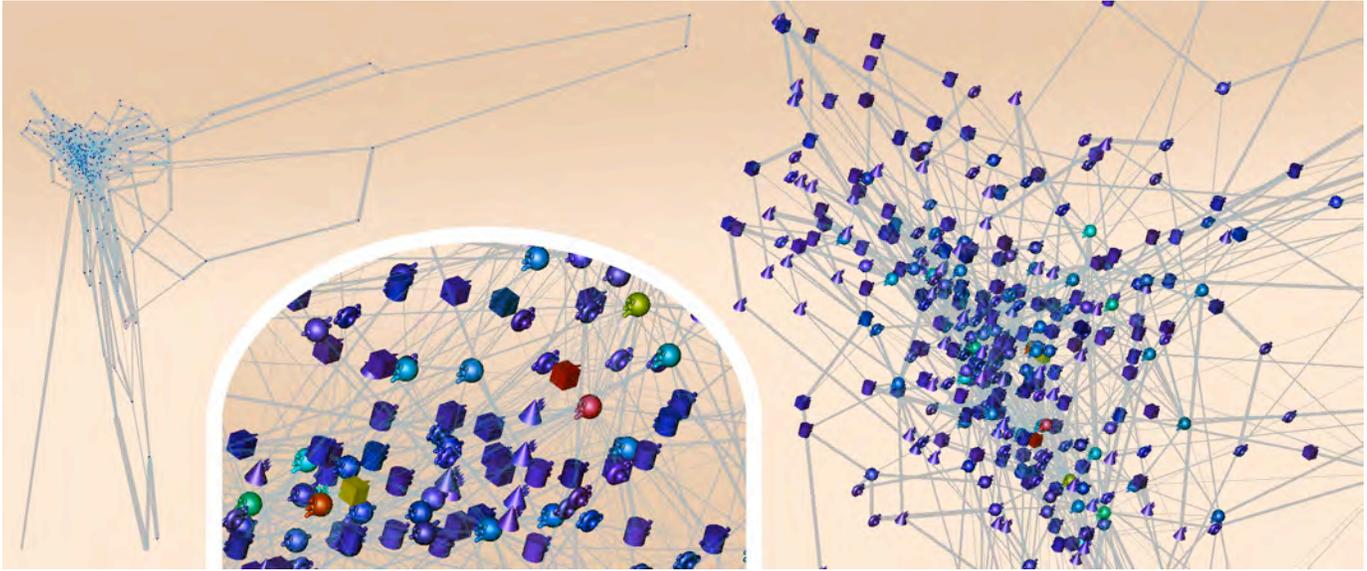

Fig. 7. VAST 2008 competition data, call connectivity subset. 3D graph was generated using Eigen state reduction from created adjacency sparse matrix. Line width signifies length of call between connected glyphs. Glyph color marks total number of minutes called by. Left arms indicate number of incoming calls. Right arms show outgoing calls. Notice set of spheres and two cubes in zoomed region in center, that have many connections, have lots of minute usage, but mostly receive calls.

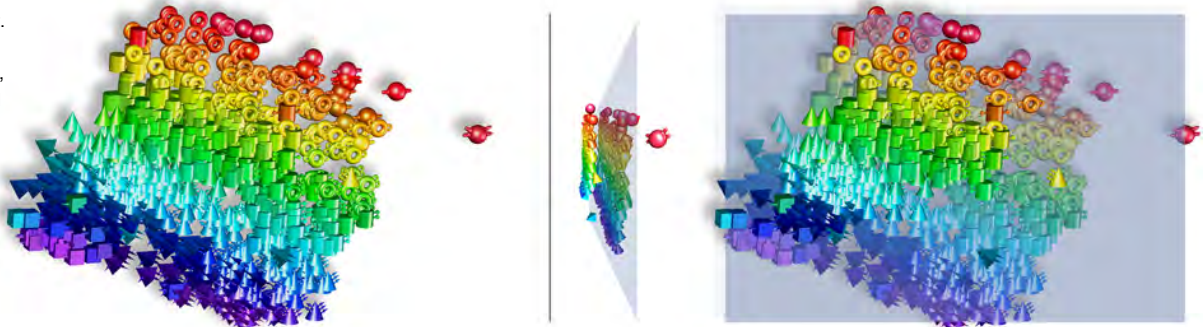

Fig. 8. Small Stars and Galaxy data set. After principal component analysis, components 2 and 1 mapped to x,y. Hue mapped to "gmag". Shape mapped to "zmag". Left arms mapped to "umag", right branches to "rmag". On left flat plot, on right plot classification with galaxies separated from stars by depth.

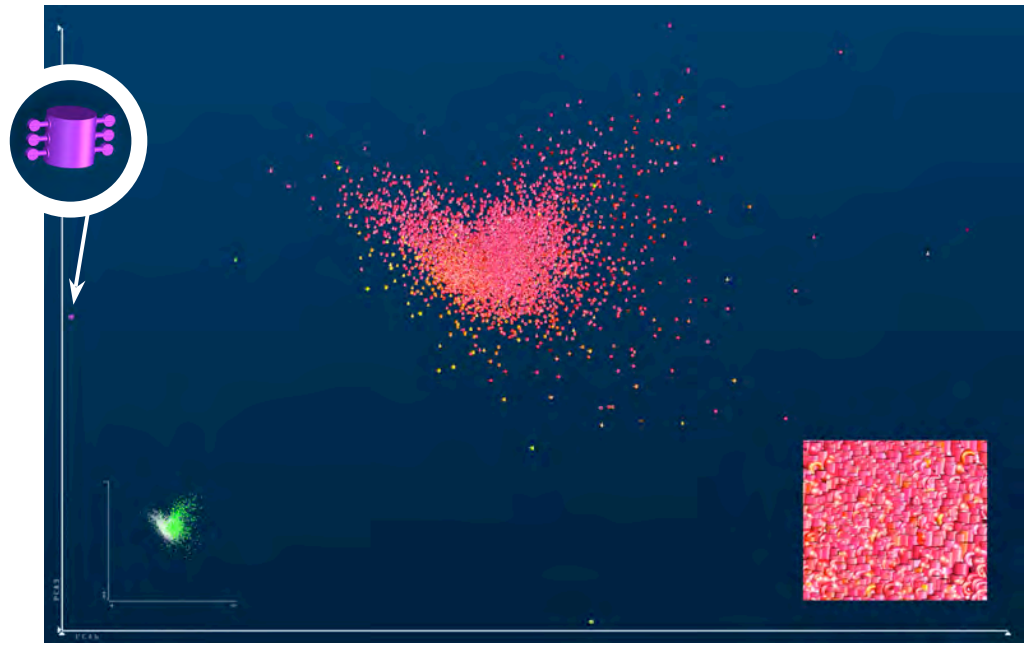

Fig. 9. Expanded view of High Energy Pysics Data set. Data set over 10,000 data records. Notice that up close, the overlap of the glyphs create a distinct texture, in which it is still possible to recognize individual shapes. Notice as well the outliers, and how even on quick inspection it is possible to recognize shape, hue, and added features (arms), which all map to specific data properties. On bottom left hint at classification between signals and background events, classification which researchers are striving for.